\documentclass[twocolumn]{aastex62}

\usepackage{graphicx}
\usepackage{txfonts}

\usepackage[capitalise]{cleveref}

\def\ms{\,m\,s$^{-1}$}         
\def\kms{\,km\,s$^{-1}$}       
\def\msol{$M_\odot$}		
\def\rsol{$R_\odot$}		
\def\denssol{$\rho_\odot$}	
\def\mstar{$M_*$}		
\def\rstar{$R_*$}		
\def\densstar{$\rho_*$}		
\def\mplanet{$M_{\rm P}$}	
\def\rplanet{$R_{\rm P}$}	
\def\mjup{$M_{\rm Jup}$}	
\def\rjup{$R_{\rm Jup}$}	

\def\teff{$T_{\rm eff}$}
\def\feh{[Fe/H]}

\def\vsini{$v_* \sin i_*$}
\def\vsinirm{$v_* \sin i_{\rm *,RM}$}
\def\vsinidt{$v_* \sin i_{\rm *,DT}$}
\def\vsinispec{$v_* \sin i_{\rm *,spec}$}
\def\mictrb{$\xi_{\rm t}$}

\def\kms{km\, s$^{-1}$}
\def\sun{$\odot$}

\newcommand{\leftcell}[1]{\multicolumn{1}{l}{#1}}

\received{September 20, 2018}

\shorttitle{A low-density hot Jupiter in a near-aligned orbit}
\shortauthors{Anderson et al.}

\begin{document}

\title{A low-density hot Jupiter in a near-aligned, 4.5-day orbit around a $V = 10.8$, F5V star}

\correspondingauthor{D. R. Anderson}
\email{d.r.anderson@keele.ac.uk}

\author[0000-0001-7416-7522]{D. R. Anderson}
\affiliation{Astrophysics Group, Keele University, Staffordshire ST5 5BG, UK}

\author{F. Bouchy}
\affiliation{Observatoire de Gen\`eve, Universit\'e de Gen\`eve, 51 Chemin 
       des Maillettes, 1290 Sauverny, Switzerland}

\author{D. J. A. Brown}
\affiliation{Department of Physics, University of Warwick, Coventry CV4 7AL, UK}
\affiliation{Centre for Exoplanets and Habitability, University of Warwick, Gibbet Hill Road, Coventry CV4 7AL, UK}

\author{A.~Burdanov}
\affiliation{Space sciences, Technologies and Astrophysics Research (STAR) Institute, Universit\'e de Li\`ege, Li\`ege 1, Belgium}

\author{A.~Collier~Cameron}
\affiliation{SUPA, School of Physics and Astronomy, University of St. Andrews, 
       North Haugh, Fife KY16 9SS, UK}

\author{L.~Delrez}
\affiliation{Space sciences, Technologies and Astrophysics Research (STAR) Institute, Universit\'e de Li\`ege, Li\`ege 1, Belgium}
\affiliation{Cavendish Laboratory, J J Thomson Avenue, Cambridge CB3 0HE, UK}

\author{M.~Gillon}
\affiliation{Space sciences, Technologies and Astrophysics Research (STAR) Institute, Universit\'e de Li\`ege, Li\`ege 1, Belgium}

\author{C.~Hellier}
\affiliation{Astrophysics Group, Keele University, Staffordshire ST5 5BG, UK}

\author{E.~Jehin}
\affiliation{Space sciences, Technologies and Astrophysics Research (STAR) Institute, Universit\'e de Li\`ege, Li\`ege 1, Belgium}

\author{M.~Lendl}
\affiliation{Observatoire de Gen\`eve, Universit\'e de Gen\`eve, 51 Chemin 
       des Maillettes, 1290 Sauverny, Switzerland}
\affiliation{Space Research Institute, Austrian Academy of Sciences, Schmiedlstr. 6, 8042 Graz, Austria}

\author{P.~F.~L.~Maxted}
\affiliation{Astrophysics Group, Keele University, Staffordshire ST5 5BG, UK}

\author{L.~D.~Nielsen}
\affiliation{Observatoire de Gen\`eve, Universit\'e de Gen\`eve, 51 Chemin 
       des Maillettes, 1290 Sauverny, Switzerland}

\author{F.~Pepe}
\affiliation{Observatoire de Gen\`eve, Universit\'e de Gen\`eve, 51 Chemin 
       des Maillettes, 1290 Sauverny, Switzerland}

\author{D.~Pollacco}
\affiliation{Department of Physics, University of Warwick, Coventry CV4 7AL, UK}
\affiliation{Centre for Exoplanets and Habitability, University of Warwick, Gibbet Hill Road, Coventry CV4 7AL, UK}

\author{D.~Queloz}
\affiliation{Observatoire de Gen\`eve, Universit\'e de Gen\`eve, 51 Chemin 
       des Maillettes, 1290 Sauverny, Switzerland}
\affiliation{Cavendish Laboratory, J J Thomson Avenue, Cambridge CB3 0HE, UK}

\author{D.~S\'egransan}
\affiliation{Observatoire de Gen\`eve, Universit\'e de Gen\`eve, 51 Chemin 
       des Maillettes, 1290 Sauverny, Switzerland}

\author{B.~Smalley}
\affiliation{Astrophysics Group, Keele University, Staffordshire ST5 5BG, UK}

\author{L.~Y.~Temple}
\affiliation{Astrophysics Group, Keele University, Staffordshire ST5 5BG, UK}

\author{A.~H.~M.~J.~Triaud}
\affiliation{School of Physics \& Astronomy, University of Birmingham, Edgbaston, Birmingham, B15 2TT, UK}

\author{S.~Udry}
\affiliation{Observatoire de Gen\`eve, Universit\'e de Gen\`eve, 51 Chemin 
       des Maillettes, 1290 Sauverny, Switzerland}

\author{R.~G.~West}
\affiliation{Department of Physics, University of Warwick, Coventry CV4 7AL, UK}
\affiliation{Centre for Exoplanets and Habitability, University of Warwick, Gibbet Hill Road, Coventry CV4 7AL, UK}



\begin{abstract}
We report the independent discovery and characterisation of a hot Jupiter in a 4.5-d, transiting orbit around the star TYC\,7282-1298-1 ($V = 10.8$, F5V). The planet has been pursued by the NGTS team as NGTS-2b and by ourselves as WASP-179b. 
We characterised the system using a combination of photometry from WASP-South and TRAPPIST-South, and spectra from CORALIE (around the orbit) and HARPS (through the transit). 
We find the planet's orbit to be nearly aligned with its star's spin. From a detection of the Rossiter-McLaughlin effect, we measure a projected stellar obliquity of $\lambda = -19 \pm 6^\circ$. 
From line-profile tomography of the same spectra, we measure $\lambda = -11 \pm 5^\circ$. 
We find the planet to have a low density (\mplanet\ = 0.67 $\pm$ 0.09\,\mjup, \rplanet\ = 1.54 $\pm$ 0.06\,\rjup), which, along with its moderately bright host star, makes it a good target for transmission spectroscopy. 
We find a lower stellar mass (\mstar\ = $1.30 \pm 0.07$\,\msol) than reported by the NGTS team (\mstar\ = $1.64 \pm 0.21$\,\msol), though the difference is only $1.5$\,$\sigma$.
\end{abstract}

\keywords{planets and satellites: individual (WASP-179b, NGTS-2b) --- stars: individual (TYC\,7282-1298-1, WASP-179, NGTS-2)}

\section{Introduction} \label{sec:intro}
Ground-based transit surveys are well matched to finding `hot Jupiters', gas-giant planets in close orbits of a few days. Such systems are among the best targets for characterisation of planetary atmospheres, particularly when the host star is relatively bright and when the planet is bloated (e.g. \citealt{2017A&A...602A..36W,2018ApJ...858L...6K,2018Natur.557...68S}). 

\citet{2018arXiv180510449R}, hereafter R18, recently reported the discovery and characterisation of a hot Jupiter in a transiting orbit around the star TYC\,7282-1298-1. 
Having detected transits of the star using the Next Generation Transit Survey (NGTS; \citealt{2018MNRAS.475.4476W}), R18 confirmed the existence of the planet and derived the system's parameters using a combination of the transit lightcurves from NGTS and radial velocities around the orbit calculated from HARPS spectra \citep{2002Msngr.110....9P}. 
R18 reported NGTS-2b to be a low-density planet (\mplanet\ = 0.74 $\pm$ 0.13\,\mjup, \rplanet\ = 1.595 $\pm$ 0.046\,\rjup) in a 4.51-d orbit around a rapidly rotating (\vsini\ = 15.2 $\pm$ 0.8\,\kms) F5V star. 

The WASP transit survey \citep{2006PASP..118.1407P} had also been following the star since finding a transit signal and adopting it as a candidate in 2013. We present here an independent discovery and characterisation of the system, which we also designate WASP-179 (noting that many planetary systems have been given designations by more than one transit team; e.g. \citealt{2018arXiv180307559L}). 
Being rare, the reporting of such an independent discovery is an important verification of the planet and a validation of the techniques of the respective surveys.

In addition to transit photometry and spectra around the orbit, we report spectra taken through a transit of the planet, from which we derive the projected stellar obliquity (projected spin-orbit angle, $\lambda$). 
Stellar obliquity is considered a diagnostic for the mechanisms via which hot Jupiters migrated to their current orbits (see \citealt{2018arXiv180106117D} for a recent review on the origins of hot Jupiters).

\section{Observations} \label{sec:obs}
\begin{figure}
\includegraphics[width=90mm]{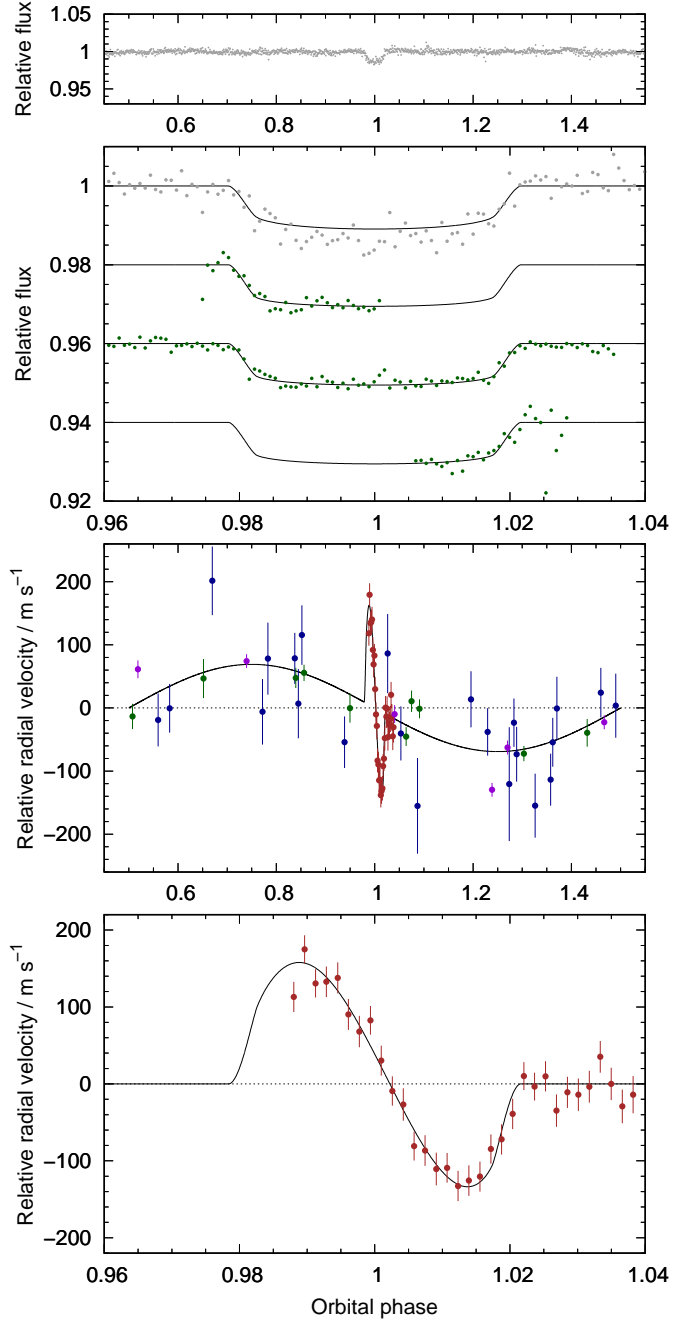}
\caption{WASP-179b discovery data. 
{\it Top panel}: WASP-South lightcurve folded on the transit ephemeris. 
{\it Second panel}: Transit lightcurves from WASP-South (grey) and TRAPPIST (green), offset for clarity, binned with a bin width equivalent to five minutes, and plotted chronologically with the most recent at the bottom. 
The best-fitting transit model is superimposed. 
{\it Third panel}: The CORALIE (blue) and HARPS (HAM = green, EGGS = violet, RM = brown) RVs with the best-fitting orbital and RM models. 
{\it Bottom panel}: The apparent radial-velocity anomaly illustrated by HARPS RVs during transit, together with the best-fitting RM effect model. The best-fitting orbital model has been subtracted.  
\label{fig:rv-phot}}
\end{figure}

We observed WASP-179 ($V = 10.8$) during 2006--2008 and during 2011--2014 with the WASP-South facility (\cref{fig:rv-phot}, top panel; \citealt{2006PASP..118.1407P}), and identified it as a candidate transiting-planet system using the techniques described in \citet{2006MNRAS.373..799C,2007MNRAS.380.1230C}.
We conducted photometric follow-up using the 0.6-m TRAPPIST-South imager \citep{2011EPJWC..1106002G,2011Msngr.145....2J} and spectroscopic follow-up using both CORALIE on the 1.2-m {\it Euler}-Swiss telescope \citep{2000A&A...354...99Q} and HARPS on the 3.6-m ESO telescope \citep{2002Msngr.110....9P}. 

From the TRAPPIST photometry we confirmed that the periodic photometric dip is on-target and that the transit ephemeris and shape are consistent with the WASP-South photometry (\cref{fig:rv-phot}, second panel). 
An unresolved star can impact the determination of the system parameters by diluting the transits (e.g. \citealt{2016ApJ...833L..19E,2018MNRAS.478.4720G}), but the Gaia DR2 \citep{2018A&A...616A...1G} excludes nearby sources beyond its angular resolution limit of 0.4\arcsec.
We computed radial-velocity (RV) measurements from the CORALIE and HARPS spectra by weighted cross-correlation with a G2 binary mask \citep{1996A&AS..119..373B,2002Msngr.110....9P}.
We detected a sinusoidal variation in the CORALIE RVs that phases with the WASP ephemeris and which has a semi-amplitude consistent with a planetary mass companion (\cref{fig:rv-phot}, third panel). 
We timed the HARPS observations to coincide with a transit, aiming to measure the projected stellar obliquity via the Rossiter-McLaughlin (RM) effect (e.g. \citealt{2012ApJ...757...18A}). 
Using an exposure time of 10\,min, we took 32 spectra in high-accuracy mode (HAM) through the transit on the night of 2018 Mar 27 (\cref{fig:rv-phot}, bottom panel). The sequence began after ingress as the telescope was previously occupied by technical intervention. 
To more precisely measure the orbital eccentricity and the amplitude of the stellar reflex motion we included in our analysis the 16 HARPS RVs from \citet{2018arXiv180510449R}: ten spectra were obtained in HAM (four with 20-min exposures and six with 40-min exposures) and six spectra were obtained in high-efficiency mode (EGGS) with 20-min exposures.
See \cref{tab:obs} for a summary of the observations used in this paper. See \cref{tab:phot} and \cref{tab:rv} for the RVs and photometry, respectively.\\

\begin{deluxetable}{lcrlc}
\tablecaption{Summary of observations \label{tab:obs}}
\tablehead{
  \colhead{Facility} & \colhead{Date\tablenotemark{a}} & \colhead{$N_{\rm obs}$} & \colhead{Notes\tablenotemark{b}}
}
\startdata
{\it Photometry}\\
WASP-South	    	& 2006 May--2014 Aug     & 77\,549 & 400--700 nm\\
TRAPPIST-South		& 2015 Apr 23			& 538	& $I+z$			\\
TRAPPIST-South		& 2018 Apr 14			& 1\,366	& $I+z$			\\
TRAPPIST-South		& 2018 Apr 23			& 400	& $I+z$			\\
\\
{\it Spectroscopy}\\
Euler/CORALIE	& 2015 Mar--2017 Apr			& 23	& orbit 		\\
ESO3.6/HARPS\tablenotemark{c}	& 2017 Jul--2018 Mar  		&  16 	& orbit 		\\
ESO3.6/HARPS	& 2018 Mar 27  		&  32 	& transit 		\\
\enddata
\tablenotetext{a}{The dates are `night beginning'.}
\tablenotetext{b}{For the photometry datasets, we state which filter was used. For the spectroscopy datasets, we indicate whether the data cover the orbit or the transit.}
\tablenotetext{c}{From \citet{2018arXiv180510449R}.}
\end{deluxetable}

\begin{figure}
\includegraphics[width=90mm]{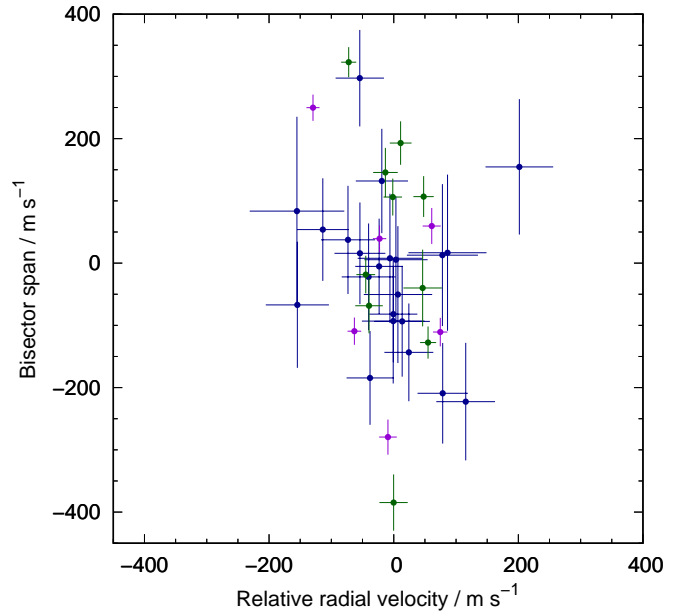}
\caption{
Bisector span versus radial velocity (CORALIE = blue symbols; HARPS HAM = green symbols; HARPS EGGS = violet symbols). 
The weak correlation is not statistically significant. 
For CORALIE: $r_{\rm weighted} = -0.29$, with weight = ${\sigma_{RV}}^{-1}$; $p$-value = 0.17. 
For HARPS: $r_{\rm weighted} = -0.40$; $p$-value = 0.12.
Combined: $r_{\rm weighted} = -0.36$; $p$-value = 0.03.
We omit the HARPS data taken through the transit as they are affected by the RM effect.
\label{fig:bis}}
\end{figure}

We checked for a corellation between RV and bisector span, which can indicate that an RV signal is the result of stellar activity \citep{2001A&A...379..279Q}, or that an RV signal and a transit signal are both due to a blended eclipsing binary \citep{2004ApJ...614..979T}. 
There is no significant correlation (\cref{fig:bis}).
Further, the detection of the RM effect and the nature of the trace in the Doppler tomogram (Section~\ref{sec:mcmc}) conclusively prove that the photometric and spectroscopic signals are induced by a planet \citep{2010MNRAS.407..507C, 2010ApJ...724.1108J}. 

\begin{table} 
\caption{Photometry} 
\label{tab:phot} 
\begin{tabular}{rrrlr} 
\hline 
\hline
\leftcell{BJD(UTC)} & \leftcell{Rel. flux, $F$} & \leftcell{$\sigma_{F}$} & Imager & Set \\ 
\leftcell{$-$2450000} &        &                 & & \\
\leftcell{(day)} & & & & \\
\hline
3860.389988 & 1.005491 & 0.005538 & WASP-South & 1 \\
3860.390324 & 0.994179 & 0.005506 & WASP-South & 1 \\
\leftcell{\ldots}\\
7136.476710 & 0.995092 & 0.010803 & TRAPPIST-S & 2 \\
7136.476940 & 0.990677 & 0.010697 & TRAPPIST-S & 2 \\
\leftcell{\ldots}\\
\hline
\end{tabular}
\begin{flushleft}
The flux values are differential and normalised to the out-of-transit levels. 
The uncertainties are the formal errors (i.e. they have not been rescaled).
This table is available in its entirety via the CDS. 
\end{flushleft}
\end{table}

\begin{table}
\caption{Radial velocities} 
\label{tab:rv} 
\begin{tabular}{rrrrl} 
\hline 
\hline
 \leftcell{BJD(UTC)} & \leftcell{RV} & \leftcell{$\sigma_{\rm RV}$} & \leftcell{BS} & \leftcell{Spectrograph}\\ 
\leftcell{$-$2450000}&    &                     &   & \\
\leftcell{(day)}     & \leftcell{(km s$^{-1}$)} & \leftcell{(km s$^{-1}$)} & \leftcell{(km s$^{-1}$)} & \\ 
\hline
7111.734763 & $-$26.3518 & 0.0508 & 0.0055 & CORALIE\\
7175.700117 & $-$26.1538 & 0.0544 & 0.1546 & CORALIE\\
\leftcell{\ldots}\\
8205.671957 & $-$26.2426 & 0.0195 & $-$0.0268 & HARPS\\
8205.679191 & $-$26.1815 & 0.0183 & $-$0.1014 & HARPS\\
\leftcell{\ldots}\\
\hline
\end{tabular}\\
Uncertainties are the formal errors (i.e. with no added jitter). 
The uncertainty on bisector span (BS) is 2\,$\sigma_{\rm RV}$. 
This table is available in its entirety via the CDS. 
\end{table}

\section{Stellar analysis} \label{sec:star}
We co-added the individual HARPS spectra from the night of 2018 Mar 27 to obtain an average signal-to-noise of 150:1. 
We performed a spectral analysis using the procedures detailed in \cite{2013MNRAS.428.3164D} to obtain stellar effective temperature, surface gravity, metallicity, and projected rotation speed. 
We calculated macroturbulence using a slight extrapolation of the calibration of \citet{2014MNRAS.444.3592D} and we calculated microturbulence using the calibration of \citet{2012MNRAS.423..122B}. 
We do not detect lithium in the spectra. 
The results of the spectral analysis are given in \cref{tab:spec}. 

We searched the WASP-South lightcurves for modulation, as may result from the combination of stellar rotation and magnetic activity, using the method of \citet{2011PASP..123..547M}. We found no convincing signal, unsurprising for an F5V star, and place an upper limit of $\sim$1\,mmag on the amplitude of any sinusoidal signal. 

\begin{deluxetable}{lccc}
\tablecaption{Stellar parameters \label{tab:spec}} 
\tablehead{
  \colhead{Parameter} & \colhead{Symbol} & \colhead{Value} & \colhead{Unit}
}
\startdata
Constellation & \ldots & Centaurus & \ldots \\
Right Ascension (J2000) & \ldots & $\rm 14^{h} 20^{m} 29\fs49$ & \ldots \\
Declination (J2000)		& \ldots & $\rm -31\degr 12\arcmin 07\fs4$	& \ldots \\	
Tycho-2 $V_{\rm mag}$	& \ldots & 10.8	& \ldots \\
2MASS $K_{\rm mag}$	& \ldots & 9.8	& \ldots \\
Spectral type\tablenotemark{a}   & \ldots & F5V & \ldots \\
Stellar effective temperature & $T_{\rm eff}$ & 6450 $\pm$ 50 & K \\
Stellar mass & \mstar & 1.302 $\pm$ 0.034 & \msol \\
Stellar radius (IRFM) & $R_{\rm *,IRFM}$ & $1.62 \pm 0.09$ & \rsol \\
Stellar surface gravity & $\log g_{*}$ & 4.1 $\pm$ 0.1 & [cgs]\\
Stellar metallicity\tablenotemark{b} & [Fe/H] & $-0.09 \pm 0.09$ & \ldots\\
Stellar luminosity & $\log$(\mstar/\msol) & 0.691 $\pm$ 0.050 & \ldots\\
Proj. stellar rotation speed & $v \sin i_{\rm *,spec}$ & $14.8 \pm 1.0$ & \kms\\
Macroturbulence & $v_{\rm mac}$ & $6.4 \pm 0.7$ & \kms\\
Microturbulence & \mictrb & $1.6 \pm 0.1$ & \kms\\
Reddening & $E(B-V)$ & 0.068 & \ldots\\
Distance & d & $350 \pm 11$ & pc \\
Age & $\tau$ & $2.7 \pm 0.2$ & Gyr \\
\enddata
\tablenotetext{a}{
Spectral type estimated using the {\sc mkclass} spectral classification code of \citet{2014AJ....147...80G}.}
\tablenotetext{b}{Iron abundance is relative to the solar value of \protect{\citet{2009ARAA..47..481A}}.}
\end{deluxetable}

We calculated the distance to WASP-179 ($d = 350 \pm 11$\,pc) using a parallax of $2.861 \pm 0.096$\,mas, which is the Gaia DR2 parallax with the correction suggested by \citet{2018ApJ...862...61S} applied.
We calculated the effective temperature ($T_{\rm eff,IRFM} = 6770 \pm 150$\,K) and angular diameter ($\theta = 0.043 \pm 0.002$\,mas) of the star using the infrared flux method (IRFM) of \citet{1977MNRAS.180..177B}, assuming reddening of $E(B-V) = 0.068$ from dust maps \citep{2011ApJ...737..103S}.
We thus calculated its luminosity ($\log(L/L_{\rm \odot}) = 0.691 \pm 0.050$) and its radius ($R_{\rm *,IRFM} = 1.62 \pm 0.09$\,\rsol), which is consistent with the value of R18 ($1.70 \pm 0.05$\,\rsol). 
If instead we use the non-corrected Gaia DR2 parallax ($2.779 \pm 0.063$\,mas) then we obtain: $d = 360 \pm 8$\,pc, $R_{\rm *,IRFM} = 1.67 \pm 0.09$\,\rsol, and $\log(L/L_{\rm \odot}) = 0.716 \pm 0.045$.

\begin{figure}
\includegraphics[width=90mm]{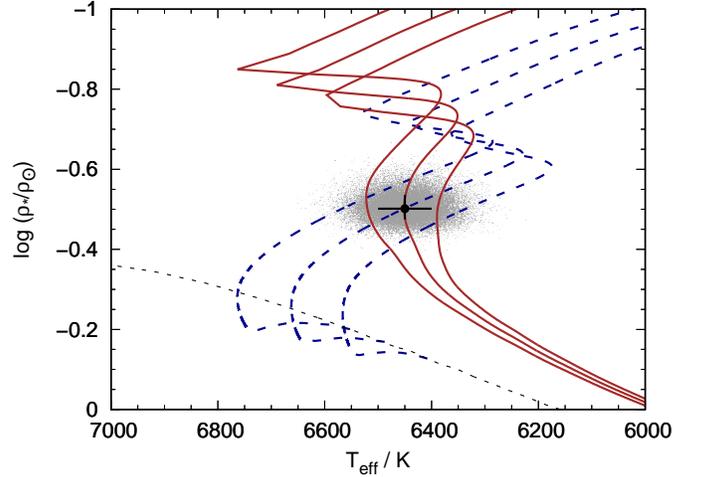}
\caption{
A modified Hertzsprung-Russel diagram showing the results of the {\sc bagemass} MCMC analysis for WASP-179. 
The grey dots are the steps in the Markov chain. 
The dotted line (black) is the ZAMS. 
The solid lines (brown) are isochrones for $\tau = 2.7 \pm 0.2$\,Gyr. 
The dashed lines (blue) are evolutionary tracks for \mstar\ = 1.302 $\pm$ 0.034\,\msol. 
The black point with error bars are the values of \teff\ and \densstar\ measured from the spectra and the transit lightcurves, respectively.
\label{fig:evol}}
\end{figure}

Though we can measure stellar density, \densstar, directly from the transit lightcurves, we require a constraint on stellar mass \mstar, or radius \rstar, for a full characterisation of the system. 
We inferred \mstar\ = 1.302 $\pm$ 0.034\,\msol\ and age $\tau$ = 2.7 $\pm$ 0.2\,Gyr using the {\sc bagemass} stellar evolution MCMC code of \citet{2015A&A...575A..36M} with input of the values of \densstar\ from an initial MCMC analysis (see Section~\ref{sec:mcmc}) and \teff\ and \feh\ from the spectral analysis (\cref{fig:evol}). 
We conservatively inflated the error bar by a factor of 2 to place a Gaussian prior on \mstar\ (1.30 $\pm$ 0.07\,\msol) in our final MCMC analysis. 
We note that we derive a value of \rstar\ ($1.62 \pm 0.06$\,\rsol) from our final MCMC that is consistent with the value that we obtained from the IRFM and Gaia parallax ($1.62 \pm 0.09$\,\rsol), which we could have used to place a Gaussian prior on \rstar\ instead.

\section{Stellar obliquity and system parameters from an MCMC analysis} 
\label{sec:mcmc}

We determined the system parameters from a simultaneous fit to the transit lightcurves and the radial velocities using the current version of the Markov-chain Monte Carlo (MCMC) code presented in \citet{2007MNRAS.380.1230C} and described further in \citet{2015A&A...575A..61A}. 
We modelled the RM effect using the formulation of \citet{2011ApJ...742...69H}. 

When we fit for an eccentric orbit we obtained $e = 0.05^{+0.04}_{-0.03}$, with a 2-$\sigma$ upper limit of $e < 0.14$. In the absence of evidence to the contrary we adopt a circular orbit, as advocated in \citet{2012MNRAS.422.1988A}. 
We accounted for stellar noise in the RV measurements by adding in quadrature with the formal RV uncertainties the level of `jitter' required to achieve $\chi^2_{\rm reduced} = 1$. 
The jitter values were: 32\,\ms\ (CORALIE RVs), 18.4\,\ms\ (HARPS HAM RVs of R18), 36.6\,\ms\ (HARPS EGGS RVs of R18). No jitter was required for our HARPS RM RVs. 
To account for instrumental and astrophysical offsets, we partitioned the four RV datasets and fit a separate systemic velocity to each of them. 
When fit separately, our CORALIE RVs suggest a slightly larger stellar reflex velocity semi-amplitude ($K_1 = 72 \pm 16$\,\ms) than do the HARPS RVs of R18 ($K_1 = 68 \pm 11$\,\ms). When we analyse all of the RVs together we get $K_1 = 69.1 \pm 8.9$\,\ms. These values are consistent both with each other and with the value of R18 ($K_1 = 65.8 \pm 9.3$\,\ms). 

\begin{figure}
\includegraphics[width=90mm]{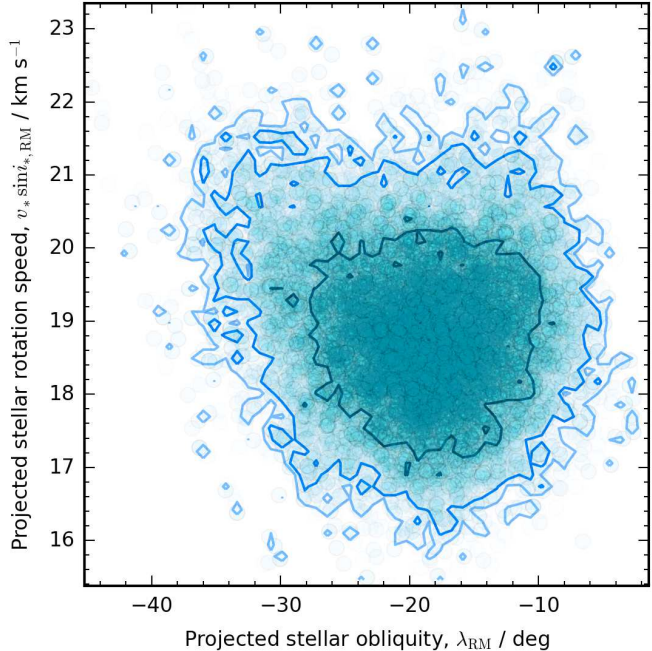}
\caption{The MCMC posterior distributions of \vsinirm\ and $\lambda_{\rm RM}$ when fitting the RM effect. The contours are the 68, 95 and 99 per cent confidence intervals. 
\label{fig:vsi-lam}}
\end{figure}

\begin{figure*}
\includegraphics[width=180mm]{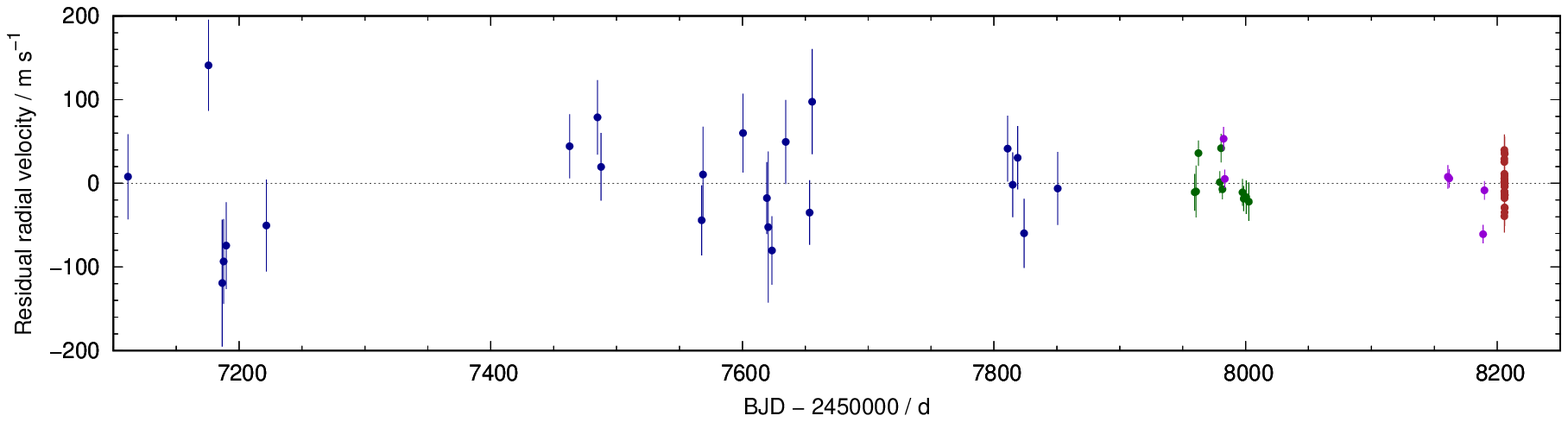}
\caption{
The residual RVs about the best-fitting Keplerian orbital and RM effect models. The symbol colours are the same as in \cref{fig:rv-phot}.
\label{fig:rv-time}}
\end{figure*}

We present the median values and 1-$\sigma$ limits on the system parameters from our final MCMC analysis in \cref{tab:mcmc}. We plot the best fits to the RVs and the transit lightcurves in \cref{fig:rv-phot}. The posterior distributions of the projected stellar rotation speed and the projected stellar obliquity indicate no degeneracy (\cref{fig:vsi-lam}), which is a result of the impact parameter being significantly non-zero ($b = 0.317 \pm 0.089$). 
We see no evidence in the RV residuals for an additional body in the system (\cref{fig:rv-time}). When we fit for a linear trend in RV, we obtained $\dot{\gamma} = -10 \pm 17$\,m\,s$^{-1}$\,yr$^{-1}$.\footnote{We excluded the RVs from the transit night, except for the final two RVs, which we fit together with the 10 HARPS HAM RVs of R18.}

\begin{figure*}
\centering
\includegraphics[width=180mm]{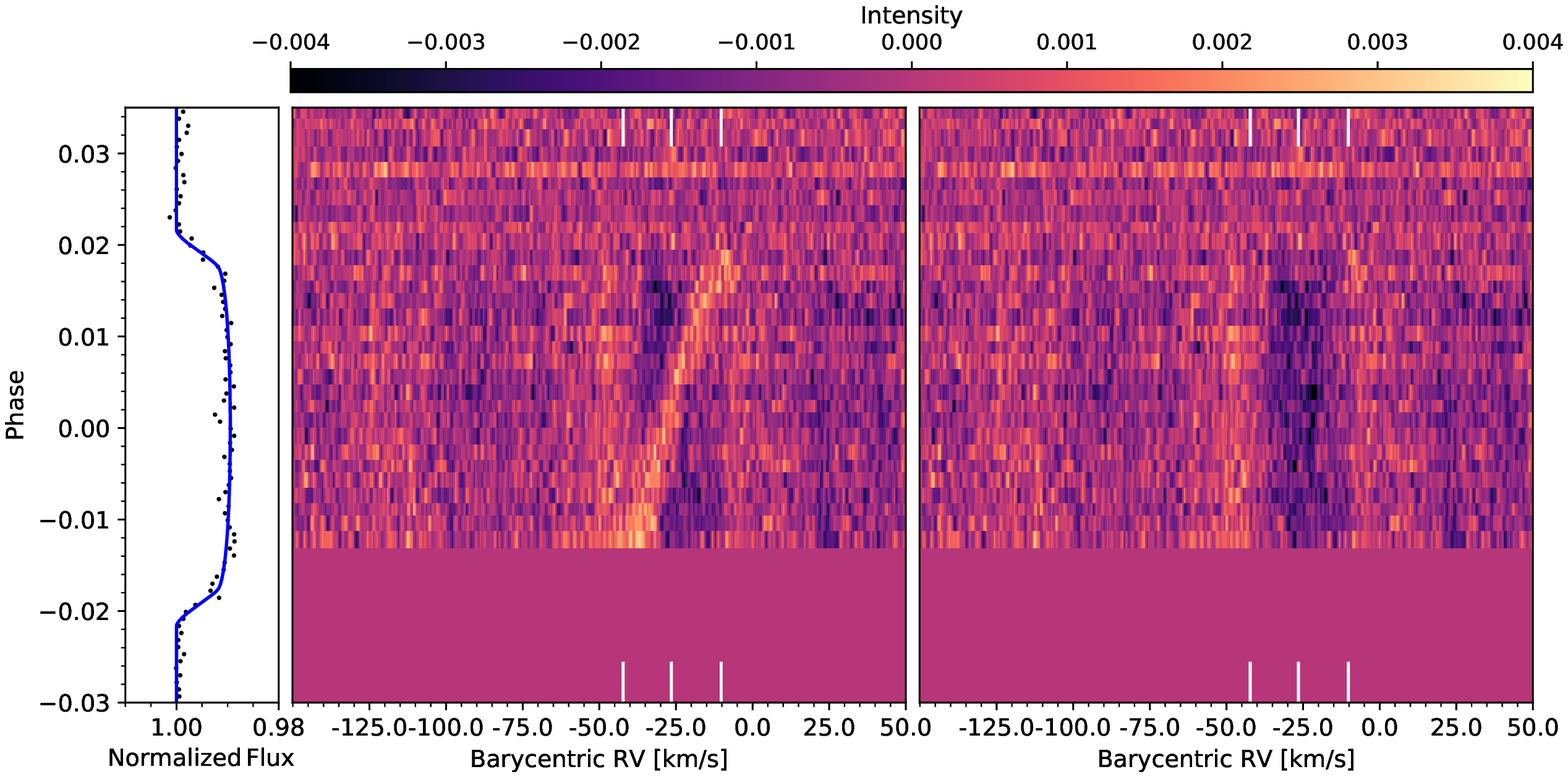}
\caption{Doppler tomogram of the HARPS spectra taken through a transit of WASP-179b. 
{\it Left panel}: The three TRAPPIST transit lightcurves combined, phase-folded on the ephemeris from \cref{tab:mcmc}, and binned with a bin width equivalent to 5\,min. This indicates the timing of the transit for comparison with the tomogram.
{\it Middle panel}: Tomogram of the residuals obtained by subtracting the average of the out-of-transit CCFs from all CCFs, leaving the bright signature of the starlight blocked by the planet during transit.
Wavelength or RV increases from left to right, time increases from bottom to top.
The white, vertical lines mark the positions of the systemic velocity ($\gamma_{\rm DT}$) and the limits of the stellar rotation speed ($\gamma_{\rm DT} \pm v_* \sin i_{\rm *,DT}$).
The stellar velocity of the planet trace moves from blue-shifted to red-shifted and covers the full range of stellar rotation, indicating a prograde, near-aligned orbit. 
{\it Right panel}: The residual tomogram after subtraction of the planet model. 
\label{fig:tomog}}
\end{figure*}

We performed an additional fit in which we measured the stellar obliquity using line-profile tomography instead of the RM effect (so we omitted the HARPS RVs taken on the transit night of 2018 Mar 27). 
We modelled the average stellar line profiles (cross-correlation functions, or CCFs) and the planet's Doppler shadow using the method presented in \citet{2010MNRAS.403..151C} and used again in \citet{2010MNRAS.407..507C}, \citet{2012ApJ...760..139B,2017MNRAS.464..810B}, and \citet{2017MNRAS.471.2743T,2018MNRAS.480.5307T}. 
We plot the tomogram (i.e. the residual map of the CCF time-series) both before and after removal of the planet model in \cref{fig:tomog}. 
The fitted parameters were the projected stellar rotation speed \vsinidt, the projected stellar obliquity $\lambda_{\rm DT}$, the impact parameter $b_{\rm DT}$, the FWHM of the line-profile perturbation due to the planet $v_{\rm FWHM}$, and the centre-of-mass velocity $\gamma_{\rm DT}$. 
We give the values of those parameters in \cref{tab:mcmc}. We omit the values of the other parameters as, depending mostly on the transit lightcurves and radial-velocity data, they are fully consistent between the two analyses. 

\begin{deluxetable*}{lccc}
\tablecaption{System parameters from an MCMC analysis\label{tab:mcmc}} 
\tablehead{
  \colhead{Parameter} & \colhead{Symbol} & \colhead{Value} & \colhead{Unit}
}
\startdata
\\
\multicolumn{4}{l}{\it MCMC Gaussian priors}\\
Stellar mass & $M_{\rm *}$ & 1.30 $\pm$ 0.07 & $M_{\rm \odot}$ \\
Stellar effective temperature & $T_{\rm eff}$ & 6450 $\pm$ 50 & K \\
\\
\multicolumn{4}{l}{\it MCMC parameters controlled by Gaussian priors}\\
Stellar mass & $M_{\rm *}$ & 1.303 $\pm$ 0.072 & $M_{\rm \odot}$\\
Stellar effective temperature & $T_{\rm eff}$ & 6453 $\pm$ 49 & K \\
\\
\multicolumn{4}{l}{\it MCMC fitted parameters}\\
Orbital period & $P$ & 4.5111204 $\pm$ 0.0000018 & d\\
Transit epoch (HJD) & $T_{\rm c}$ & 2457501.99114 $\pm$ 0.00050 & d\\
Transit duration & $T_{\rm 14}$ & 0.1944 $\pm$ 0.0016 & d\\
Planet-to-star area ratio & $R_{\rm P}^{2}$/R$_{*}^{2}$ & 0.00952 $\pm$ 0.00017 & \ldots \\
Impact parameter\tablenotemark{a} & $b$ & 0.317 $\pm$ 0.089 & \ldots \\
Reflex velocity semi-amplitude & $K_{\rm 1}$ & 69.1 $\pm$ 8.9 & \ms\\
Systemic velocity (CORALIE) & $\gamma_{\rm CORALIE}$ & $-$26\,355 $\pm$ 12 & m s$^{-1}$ \\
Systemic velocity (HARPS,RM) & $\gamma_{\rm HARPS,RM}$ & $-$26\,360.9 $\pm$ 3.5 & m s$^{-1}$ \\
Systemic velocity (HARPS,HAM) & $\gamma_{\rm HARPS,HAM}$ & $-$26\,361.7 $\pm$ 8.1 & m s$^{-1}$ \\
Systemic velocity (HARPS,EGGS) & $\gamma_{\rm HARPS,EGGS}$ & $-$26\,402 $\pm$ 15 & m s$^{-1}$ \\
Orbital eccentricity & $e$ & 0 (assumed; $<$ 0.14 at 2$\sigma$) & \ldots \\
\\
\multicolumn{4}{l}{\it MCMC derived parameters}\\
Sky-projected stellar obliquity & $\lambda_{\rm RM}$ & $-$19.0 $\pm$ 6.1 & $^\circ$ \\
Sky-projected stellar rotation speed & $v \sin i_{\rm *,RM}$ & 18.9 $\pm$ 1.1 & \kms\\
Scaled semi-major axis & $a/R_{\rm *}$ & $7.77 \pm 0.24$ & \ldots \\
Orbital inclination & $i$ & 87.66 $\pm$ 0.73 & $^\circ$\\
Ingress and egress duration & $T_{\rm 12}=T_{\rm 34}$ & 0.0190 $\pm$ 0.0013 & d\\
Stellar radius & $R_{\rm *}$ & 1.619 $\pm$ 0.058 & $R_{\rm \odot}$\\
Stellar surface gravity & $\log g_{*}$ & 4.135 $\pm$ 0.028 & [cgs]\\
Stellar density & $\rho_{\rm *}$ & 0.309 $\pm$ 0.028 & $\rho_{\rm \odot}$\\
Planetary mass & $M_{\rm P}$ & 0.670 $\pm$ 0.089 & $M_{\rm Jup}$\\
Planetary radius & $R_{\rm P}$ & 1.536 $\pm$ 0.062 & $R_{\rm Jup}$\\
Planetary surface gravity & $\log g_{\rm P}$& 2.810 $\pm$ 0.068 & [cgs]\\
Planetary density & $\rho_{\rm P}$ & 0.183 $\pm$ 0.033 & $\rho_{\rm J}$\\
Orbital semi-major axis & $a$ & 0.0584 $\pm$ 0.0011 & AU\\
Planetary equilibrium temperature\tablenotemark{b} & $T_{\rm eql}$ & 1638 $\pm$ 29 & K\\
\\
\multicolumn{4}{l}{\it Parameters from a separate MCMC including Doppler tomography}\\
Sky-projected stellar obliquity & $\lambda_{\rm DT}$ & $-$11.3 $\pm$ 4.8 & $^\circ$ \\
Sky-projected stellar rotation speed & $v \sin i_{\rm *,DT}$ & 15.91 $\pm$ 0.49 & \kms\\
Intrinsic linewidth & $v_{\rm FWHM}$ & 8.98 $\pm$ 0.32 & \kms\\
Impact parameter & $b_{\rm DT}$ & 0.206 $\pm$ 0.080 & \ldots \\
Systemic velocity & $\gamma_{\rm DT}$ & $-27\,520 \pm 400$ & \ms \\
\enddata
\tablenotetext{a}{Impact parameter is the distance between the centre of the stellar disc and the transit chord: $b = a \cos i / R_{\rm *}$.}
\tablenotetext{b}{Equilibrium temperature calculated assuming zero albedo and efficient redistribution of heat from the planet's presumed permanent day-side to its night-side.}
\end{deluxetable*}

\section{Discussion} \label{sec:disc}
We have reported the characterisation of WASP-179b (NGTS-2b), a hot Jupiter (\mplanet\ = 0.67 $\pm$ 0.09\,\mjup, \rplanet\ = 1.54 $\pm$ 0.06\,\rjup) in a 4.51-d orbit around a $V = 10.8$, F5V star. 
As a low-density planet orbiting a relatively bright star, WASP-179b is a good target for atmospheric characterisation via transmission spectroscopy (e.g. \citealt{2018Natur.557...68S}). We predict an atmospheric scale height of $\sim$1050\,km and a transmission signal similar in amplitude to that of WASP-139b \citep{2017MNRAS.465.3693H} and one tenth that of WASP-107b (see table 4 of \citealt{2017A&A...604A.110A}), which are both bloated super-Neptunes. 

From an observation of the RM effect, we find the planet to be in a prograde orbit, with a slight misalignment between the planet's orbital axis and the star's spin axis ($\lambda_{\rm RM} = -19 \pm 6^\circ$).
We find this to be corroborated by our tomographic analysis of the same transit spectra ($\lambda_{\rm RM} = -11 \pm 5^\circ$). 
The near-alignment of the system and the near-circular orbit ($e = 0.05^{+0.04}_{-0.03}$; $e < 0.14$ at 2$\sigma$) are compatible with WASP-179b having arrived in its current orbit via disc migration (see, e.g. \citealt{2018arXiv180106117D}). 
High-eccentricity migration (e.g. \citealt{2015ApJ...805...75P}) is not ruled out, however, as we calculate a circularisation timescale of just 20\,Myr (e.g. \citealt{2008ApJ...678.1396J}; assuming $Q'_P = 10^5$).
We obtain a larger estimate of the stellar rotation speed from fitting the RM effect with the Hirano model than we do from both our tomographic and spectral analyses (\vsinirm\ = $18.9 \pm 1.1$\,\kms, \vsinidt\ = $15.91 \pm 0.49$\,\kms, \vsinispec\ = $14.8 \pm 1.0$\,\kms), as was observed previously for other systems by \citet{2017MNRAS.464..810B}.

The system parameters reported by R18 differ somewhat to those presented herein. Most notably, their stellar mass (\mstar\ = $1.64 \pm 0.21$\,\msol) is a little higher than ours (\mstar\ = $1.30 \pm 0.07$\,\msol). 
Thus we find a smaller planetary mass than do R18: \mplanet\ = 0.67 $\pm$ 0.09\,\mjup\ compared to \mplanet\ = 0.74 $\pm$ 0.13\,\mjup\ (the difference is smaller than suggested by \mstar\ as we measured a larger stellar reflex velocity amplitude).
Whilst we both measured \densstar\ from our respective transit lightcurves, 
we derived \rstar\ from \densstar\ and \mstar\ (obtained from stellar models), 
whereas R18 derived \mstar\ from \densstar\ and \rstar\ (obtained from SED fitting and the Gaia parallax). 
Our values of \mstar\ ($1.30 \pm 0.07$\,\msol) and \rstar\ (1.62 $\pm$ 0.06\,\rstar) are in good agreement both with the values obtained by R18 from stellar models (\mstar\ = 1.32 $\pm$ 0.09\,\msol\ and \rstar\ = 1.58 $\pm$ 0.22\,\rsol) and the values we obtain from the empirical calibrations of \citet{2011MNRAS.417.2166S}: \mstar\ = 1.34\,\msol\ and \rstar\ = 1.63\,\rsol. 
Further, our value of \mstar\ is consistent with the F5V spectral type that we obtained using the {\sc mkclass} spectral classification code of \citet{2014AJ....147...80G}.
We note that the discrepancy is not large: the two \mstar\ values agree at the 1.5-$\sigma$ level.
This small discrepancy is somewhat due to the difference in stellar density measured from the transit lightcurves (we found \densstar\ = 0.31 $\pm$ 0.03\,\denssol\ and R18 found \densstar\ = 0.33 $\pm$ 0.05\,\denssol), but it is more so due to R18's larger stellar radius. 
R18 obtained \rstar\ = $1.70 \pm 0.05$\,\rsol\ from SED fitting and the Gaia parallax, whereas we obtained \rstar\ = $1.62 \pm 0.09$\,\rsol\ using a similar method, and we derived \rstar\ = $1.62 \pm 0.06$\,\rsol\ from our \densstar\ and \mstar\ values. 
Thus we find a smaller planetary radius than do R18: \rplanet\ = 1.54 $\pm$ 0.06\,\rjup\ compared to \rplanet\ = 1.60 $\pm$ 0.05\,\rjup\ (the difference is slightly smaller than suggested by \rstar\ as we measured a slightly larger planet-to-star area ratio).

Due to the 12-yr baseline (2006-2018) of our transit observations, our ephemeris is considerably more precise than that of R18 (8-month baseline). 
Our error bars on the orbital period and on the time of mid-transit are smaller than those of R18 by factors of 34 and 3, respectively.

\acknowledgments
WASP-South is hosted by the South African Astronomical Observatory; we are grateful for their ongoing support and assistance. 
Funding for SuperWASP comes from consortium universities and from the UK's Science and Technology Facilities Council. 
The Swiss {\it Euler} Telescope is operated by the University of Geneva, and is funded by the Swiss National Science Foundation. 
The research leading to these results has received funding from the  ARC grant for Concerted Research Actions, financed by the Wallonia-Brussels Federation. TRAPPIST is funded by the Belgian Fund for Scientific Research (Fond National de la Recherche Scientifique, FNRS) under the grant FRFC 2.5.594.09.F, with the participation of the Swiss National Science Foundation (SNF). MG and EJ are FNRS Senior Research Associates.
Based on observations collected at the European Organisation for Astronomical Research in the Southern Hemisphere under ESO programmes 099.C-0303(A), 099.C-0898(A) and 0100.C-0847(A).

\vspace{5mm}
\facilities{SuperWASP, TRAPPIST, Euler1.2m(CORALIE), ESO:3.6m(HARPS)}


\end{document}